\documentclass[a4paper,11pt]{article}
\pdfoutput=1 % if your are submitting a pdflatex (i.e. if you have
\usepackage{jcappub} % for details on the use of the package, please
\usepackage[T1]{fontenc} % if needed

\def\lsim{\mathrel{\raise.3ex\hbox{$<$\kern-.75em\lower1ex\hbox{$\sim$}}}}
\def\gsim{\mathrel{\raise.3ex\hbox{$>$\kern-.75em\lower1ex\hbox{$\sim$}}}}

\proceeding{\hfill FERMILAB-PUB-20-352-AE-T}

\title{\boldmath Revisiting AGN as the Source of IceCube's Diffuse Neutrino Flux}

\author[a,b]{Daniel Smith,}
\author[a,c,e]{Dan Hooper}
\author[a,b,d]{and Abigail Vieregg}

\affiliation[a]{University of Chicago, Kavli Institute for Cosmological Physics, Chicago, IL 60637}
\affiliation[b]{University of Chicago, Department of Physics, Chicago, IL 60637}
\affiliation[c]{University of Chicago, Department of Astronomy and Astrophysics, Chicago, IL 60637}
\affiliation[d]{University of Chicago, Enrico Fermi Institute, Chicago, IL 60637}
\affiliation[e]{Fermi National Accelerator Laboratory, Theoretical Astrophysics, Batavia, IL 60510}
\emailAdd{danielsmith@uchicago.edu}
\emailAdd{dhooper@fnal.gov}
\emailAdd{avieregg@kicp.uchicago.edu}

\abstract{The origin of the astrophysical neutrino flux reported by the IceCube Collaboration remains an open question. In this study, we use three years of publicly available IceCube data to search for evidence of neutrino emission from the blazars and non-blazar Active Galactic Nuclei (AGN) contained the Fermi 4LAC catalog. We find no evidence that these sources produce high-energy neutrinos, and conclude that blazars can produce no more than 15\% of IceCube's observed flux. The constraint we derive on the contribution from non-blazar AGN, which are less luminous and more numerous than blazars, is significantly less restrictive, and it remains possible that this class of sources could produce the entirety of the diffuse neutrino flux observed by IceCube. We anticipate that it will become possible to definitively test such scenarios as IceCube accumulates and releases more data, and as gamma-ray catalogs of AGN become increasingly complete. We also comment on starburst and other starforming galaxies, and conclude that these sources could contribute substantially to the signal observed by IceCube, in particular at the lowest detected energies.}

\begin{document}
\maketitle
\flushbottom

\section{Introduction}

In 2013, the IceCube Collaboration reported the detection of a diffuse flux of astrophysical high-energy neutrinos. The spectrum of these neutrinos is consistent with a power-law with an index of $ \alpha \approx 2.4-2.6$, extending from tens of TeV to several PeV \cite{Aartsen:2016xlq}, and with flavor ratios that are consistent with those predicted from the decays of charged pions~\cite{Aartsen:2015ivb}. The angular distribution of this flux shows no significant departures from isotropy, and searches for individual point sources in the IceCube data have thus far not resulted in any detections~\cite{Aartsen:2019fau, Aartsen:2016oji, Aartsen:2014cva} (with the possible exception of the blazar TXS 0506+056~\cite{IceCube:2018dnn,IceCube:2018cha}). These results indicate that IceCube's high-energy neutrinos are produced by a large number of extragalactic sources, of which even the brightest contribute only a small fraction of the total flux. 

Many varieties of astrophysical objects have been proposed as potential sources of high-energy neutrinos, including gamma-ray bursts~\cite{Waxman:1997ti,Piran:1999kx,Vietri:1998nm,Meszaros:2001ms,Guetta:2003wi}, star-forming and starburst galaxies~\cite{Loeb:2006tw, Lunardini:2019zcf}, both blazar and non-blazar active galactic nuclei (AGN)~\cite{Stecker:1991vm,Halzen:1997hw,Atoyan:2001ey,Mannheim:1995mm,Muecke:2002bi}, as well as tidal disruption events~\cite{Wang:2015mmh,Senno:2016bso,Dai:2016gtz,Lunardini:2016xwi,Murase:2020lnu}, and fast radio bursts~\cite{Kheirandish:2019dii,Aartsen:2017zvw,Fahey:2016czk}. Most of these proposals, however, have since been excluded as the primary source of IceCube's diffuse flux. In particular, the lack of neutrino events observed in coincidence with known gamma-ray bursts has ruled out this class of objects as a major source of IceCube's neutrinos~\cite{Aartsen:2016qcr,2012Natur.484..351I,Aartsen:2018fpd} (with the possible exception of low-luminosity gamma-ray bursts~\cite{Murase:2013ffa,Tamborra:2015qza,Senno:2015tsn}). Similarly, the lack of neutrino events from the direction of gamma-ray blazars excludes this class of sources~\cite{Aartsen:2016lir,Hooper:2018wyk}. Starburst and other starforming galaxies are also unable to generate the entirety of this signal without simultaneously exceeding the measured intensity of the isotropic gamma-ray background~\cite{Bechtol:2015uqb}.

In this paper, we attempt to measure or constrain the fraction of IceCube's diffuse neutrino flux that originates from blazars and non-blazar AGN. We also consider starburst galaxies and other starforming galaxies within this context. To this end, we follow a procedure similar to that taken in Refs.~\cite{Aartsen:2014cva,Aartsen:2016lir,Hooper:2018wyk} in searching for correlations between the arrival directions of IceCube's muon track events and the locations of known AGN. We improve and expand upon these previous studies by utilizing the latest (3 year) public release of IceCube muon track data~\cite{IceCube:Dataset}, as well as the latest catalog of gamma-ray emitting AGN detected by the Fermi Gamma-Ray Space Telescope  (the 4LAC catalog)~\cite{Fermi-LAT:2019pir}. The latter update is particularly important in the case of non-blazar AGN, as the current catalog of such sources detected by Fermi is much larger than those used in previous studies (65, compared to 19 non-blazar AGN used in Ref.~\cite{Hooper:2018wyk}, for example). Despite these advances, we do not identify any statistically significant correlation between IceCube events and known blazars or non-blazar AGN. From this, we are able to conclude that no more than 15\% of IceCube's diffuse high-energy neutrino flux can originate from blazars. The constraint we derive on neutrinos from non-blazar AGN is significantly less restrictive, and it remains possible that this class of sources could produce the entirety of the astrophysical flux observed by IceCube. This scenario is expected to be within the reach of IceCube after relatively modest increases in exposure, and with expansions of gamma-ray catalogs of non-blazar AGN. This is particularly interesting given the evidence that non-blazar AGN produce the majority of the isotropic gamma-ray background~\cite{Hooper:2016gjy}, and are likely to represent the dominant source of IceCube's diffuse neutrino flux~\cite{Hooper:2016jls}. Starburst and other starforming galaxies may also substantially contribute to the signal observed by IceCube, in particular at the lowest detected energies.

\section{Methods}
\label{sect:methods}

In this analysis, we make use of IceCube's most recent public data release of muon tracks, consisting of events collected between June 2010 and May 2013~\cite{IceCube:Dataset}.\footnote{https://icecube.wisc.edu/science/data/PS-3years} The first year of this data is from the 79-string detector, while the remaining two years are from the complete, 86-string detector. Each year of data contains 93133, 136244, and 105300 events, respectively. Each event has a reported direction, angular resolution, and a quantity known as the ``energy proxy'', which is related to the energy deposited in the detector. Each year of this dataset includes an effective area for the detector as determined by simulation, as a function of declination and neutrino energy.

To test for evidence of a neutrino signal from an individual point source, we follow the approach outlined in Ref.~\cite{Braun_2010}. The likelihood that a given source results in $n_s$ events, out of a total $N$ recorded in the detector, is given by:
\begin{equation}
    \mathcal{L}(n_s) = \prod_i^N\left[ 
    \frac{n_s}{N} S_i(|\vec{x}_s - \vec{x}_i|) + \left( 1 - \frac{n_s}{N} \right)B_i(\sin \delta_i)
    \right],
    \label{eq:prob}
\end{equation}

\noindent where $S_i$ and $B_i$ are the signal and background probability distribution functions (PDFs), respectively. These PDFs are defined as follows:
\begin{eqnarray}
    S_i &=& \frac{1}{2 \pi \sigma_i^2} \, e^{-\frac{|\vec{x}_s - \vec{x}_i|^2}{2 \sigma_i^2}} \\
    B_i &=& \frac{\mathcal{P}_B(\sin \delta_i)}{2 \pi}, \nonumber
\end{eqnarray}
where $\vec{x}_s$ is the direction to the source, $\vec{x}_i$ is the reported direction of the event, and $\sigma_i$ is the angular resolution of the event. The function, $P_B$, is equal to the fraction of events in the dataset averaged across a band of $\pm 6^\circ$ in declination, $\delta$, around a given source. We only consider sources with declination between $\pm87^\circ$ due to the limited amount of solid angle near the poles with which to characterize the background PDF.

The statistical significance in favor of a neutrino point source over a background-only hypothesis is calculated using the following test statistic:
\begin{equation}
    2 \Delta \mathcal{L}(n_s) = 2 \big[\ln \mathcal{L}(n_s) - \ln \mathcal{L}(0)\big],
    \label{eq:likelihood}
\end{equation}
where $\mathcal{L}(0)$ is the likelihood for the background-only hypothesis. From this, the $p$-value can be calculated by performing an integral over a $\chi^2$ distribution with one degree-of-freedom. 

\begin{figure}[t]
\centering
\includegraphics[width=0.8\textwidth]{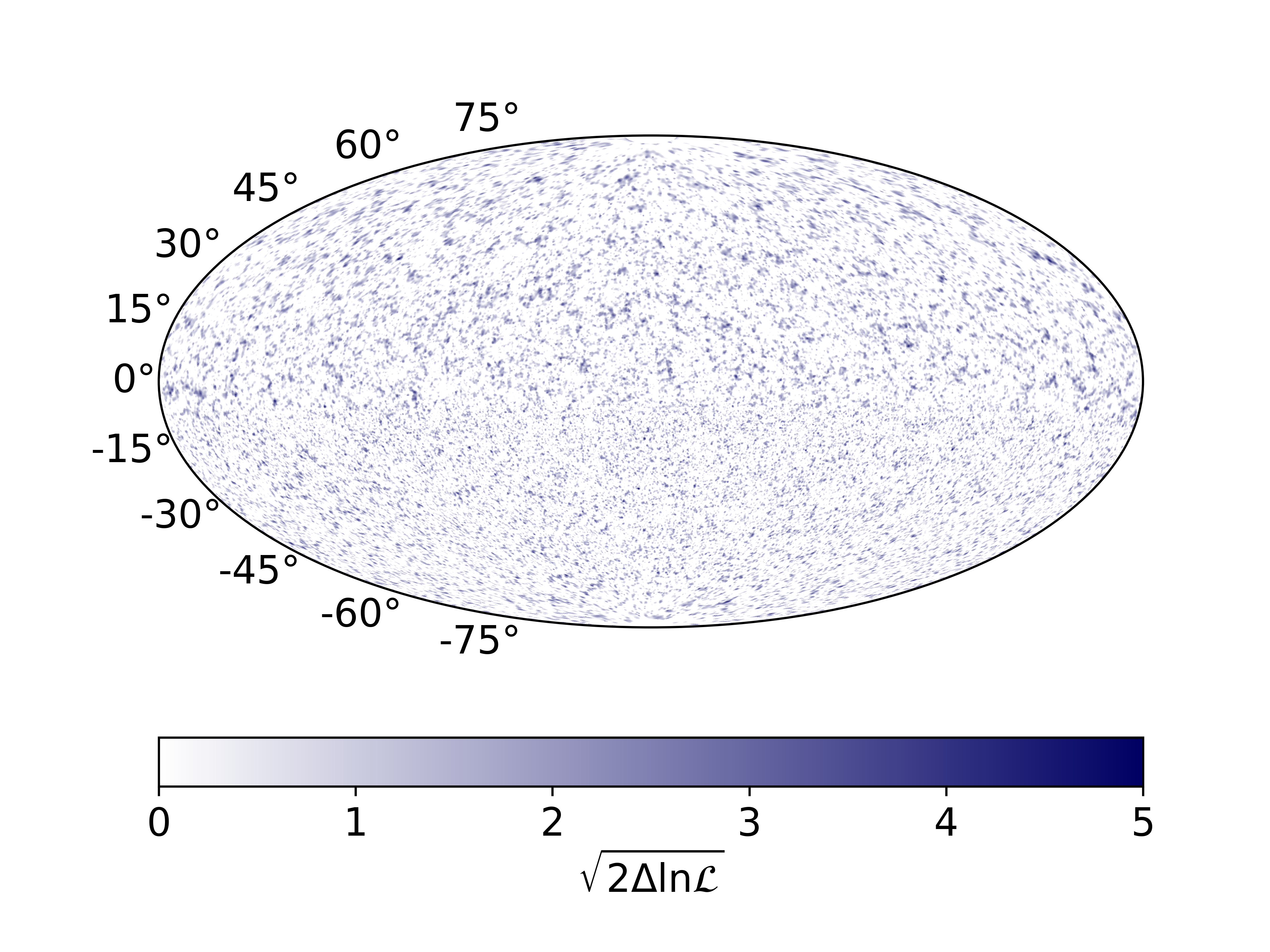}
\caption{An all-sky map of the likelihood of a neutrino point source, $\sqrt{2 \Delta \ln \mathcal{L}}$, in RA and Dec in Aitoff projection.} 
\label{fig:skymap}
\end{figure}

In contrast to the analysis performed by the IceCube Collaboration~\cite{Aartsen:2016oji}, we do not include energy information in our analysis, due to our inability to reliably relate the ``energy proxy'' provided in the public dataset with the actual energy of the neutrinos without access to the full IceCube detector simulation. The energy enters our analysis only in the form of IceCube's energy-dependent effective area. This analysis also differs from that of the IceCube Collaboration in that their analysis uses 7 years of data, while our study is limited to the 3 years of data that is publicly available at this time. 

\begin{figure}[!h]
\centering
\includegraphics[width=0.6\textwidth]{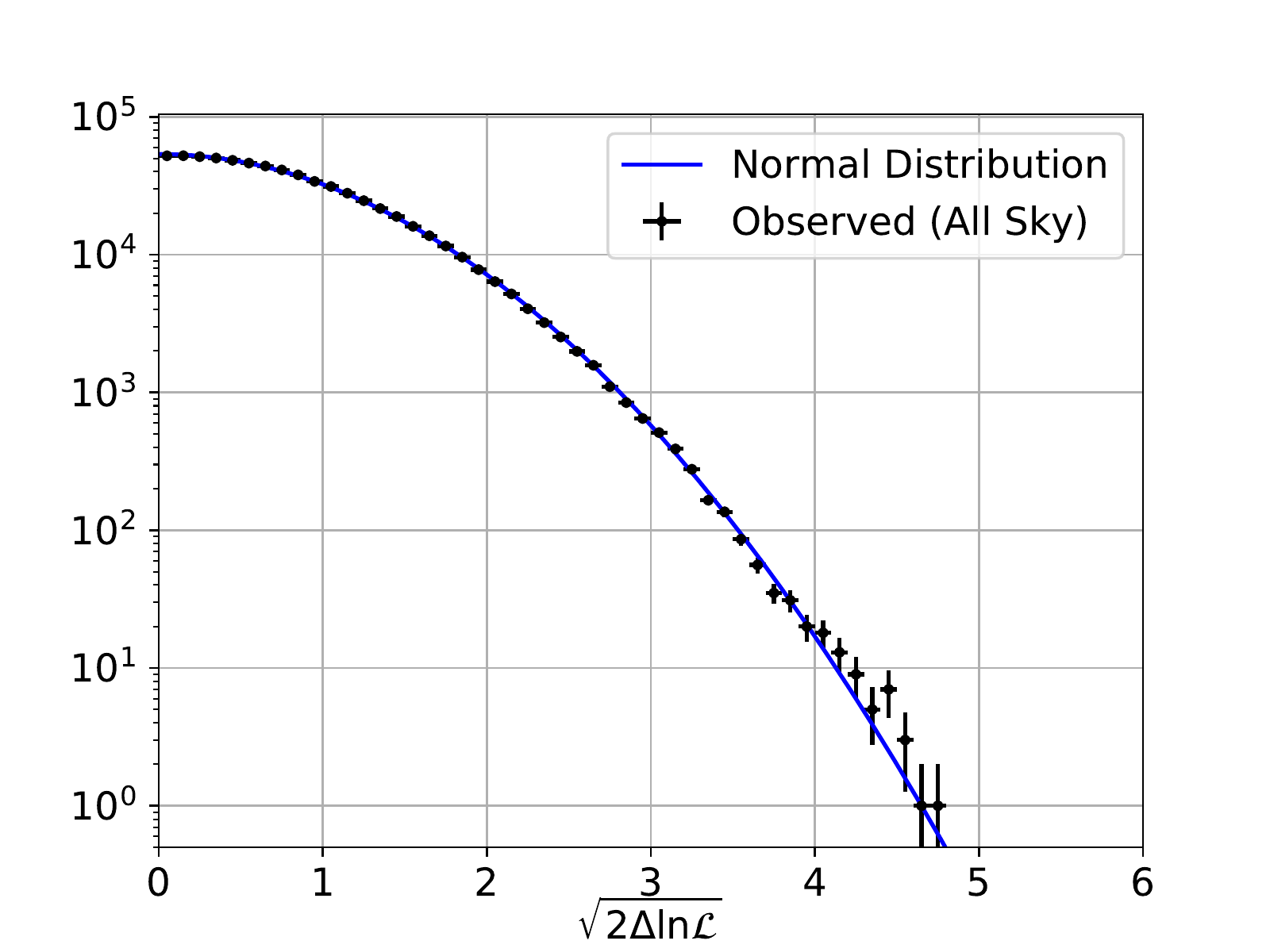}
\caption{ The likelihood distribution in favor of a neutrino point source from our all-sky scan. The observed distribution is consistent with background and we identify no evidence of a neutrino point source population. Sky locations with $\Delta \ln \mathcal{L} < 0$, corresponding to a best fit with a negative point source flux, are not shown. The error bars represent the 68\% Poissonian confidence interval on each bin.} 
\label{fig:skyhistogram}
\end{figure}

In Fig.~\ref{fig:skymap}, we show an all-sky map of the likelihood of a neutrino point source, $\sqrt{2\Delta \ln \mathcal{L}}$, in terms of right ascension and declination in Aitoff projection. This scan was performed in steps of 0.1$^\circ$, and at each point we show the result using the value of $n_s$ that maximizes the test statistic defined in Eq.~\ref{eq:likelihood}. In Fig.~\ref{fig:skyhistogram}, we show that the distribution of this test statistic across the sky is consistent with that expected from Gaussian fluctuations. As in Ref.~\cite{Hooper:2018wyk}, we identify a slight excess of events in the range of $\sqrt{2 \Delta \ln \mathcal{L}} \approx 4 - 5$. Given that this could quite plausibly be attributed to inaccuracies in the background PDF, we do not consider this to be evidence of a near-threshold point source population. In Table~\ref{tbl:fitmap}, we list the six most significant points identified in our all-sky scan. After accounting for an appropriate trials factor, these do not represent statistically significant point sources.

\begin{table}[!h]
\centering
 \begin{tabular}{|c | c | c | c | } 
 \hline
 $2 \Delta \ln \mathcal{L}$ & Pre-Trial $p$-value & RA & Dec \\ [0.5ex] 
 \hline\hline
22.28 & $2.36 \times 10^{-6}$ & 174.6 & -39.2 \\
\hline
21.72 & $3.15 \times 10^{-6}$ & 296.4 & -21.0 \\
\hline
20.79 & $5.12 \times 10^{-6}$ & 67.8 & 38.8 \\
\hline
20.25 & $6.79 \times 10^{-6}$ & 349.0 & 31.0 \\
\hline
19.80 & $8.60 \times 10^{-6}$ & 275.6 & 11.4 \\
\hline
19.36 & $1.08 \times 10^{-5}$ & 42.0 & -5.6 \\
\hline
\end{tabular}
\caption{The six most significant independent locations identified by our all-sky scan. After accounting for an appropriate trial factor, these do not represent statistically significant point sources.}
\label{tbl:fitmap}
\end{table}

%%%%%%%%%%%%%%

Next, we use the public IceCube dataset to measure or constrain the fraction of IceCube's diffuse neutrino flux that originates from known classes of astrophysical objects. The joint likelihood in each case is calculated as the product of the likelihoods for each source, as described in Eq.~\ref{eq:prob}, as a function of the total neutrino flux from the entire source population. 

In calculating the joint probability, we consider three different hypotheses for the expected neutrino fluxes from the members of a given source class:
\begin{enumerate}
\item Gamma-Ray Scaling: The neutrino flux from a given source is taken to be proportional to the gamma-ray flux observed from that source, as reported in the 4LAC catalog~\cite{Fermi-LAT:2019pir}. Here we take the gamma-ray flux to be the number of photons between 1-100 GeV per area, per time.  Such a proportionality would be expected if the observed gamma-ray emission is produced mostly through hadronic interactions, resulting in a fixed ratio of neutrinos and photons.
\item Geometrical Scaling: The neutrino flux from a given source is taken to be proportional to $1/D_L^2$, where $D_L$ is the luminosity distance of the source. This hypothesis treats the neutrino luminosity of a given source as uncorrelated to other information, and only takes into account the distance from the source. Since this approach can only make use of sources with a measured redshift, it sometimes requires the use of smaller source catalogs.
\item Flat Scaling: The neutrino flux from a given source is taken to be entirely uncorrelated to any other information under consideration. This hypothesis is maximally conservative, in that it is predicted to produce a constraint that is valid for any distribution of neutrino fluxes from the collection of sources under consideration~\cite{Aartsen:2016lir}.

\end{enumerate}

These hypotheses were each applied in our previous study~\cite{Hooper:2018wyk}, as was the case of gamma-ray scaling in Ref.~\cite{Aartsen:2019fau}. Here, we expand upon this earlier work by considering a range of updated source catalogs and utilizing the 3-year public data release of IceCube muon track events.

\section{Searching for Neutrinos from 4LAC Blazars}
\label{sect:4lac_blazar}

In this section, we consider the blazars contained within the Fourth Catalog of AGN detected by the Fermi Large Area Telescope (the 4LAC catalog). More specifically, of the 2863 sources in this catalog, we consider the 2796 classified as blazars, which includes 658 flat spectrum radio quasars (FSRQs), 1067 BL Lacs, and 1071 other sources classified as ``blazars of unknown type''. There are three AGN in this catalog that we do not consider due to their location within $3^\circ$ of the poles (for which we are not able to reliably characterize the background distribution).

\begin{table}
\centering
 \begin{tabular}{|c | c | c | c | c |}% c | c | } 
 \hline
 4FGL Name &  AGN Type & z & $\Phi_{1-100 \,\textnormal{GeV}}$ & $2 \Delta \ln \mathcal{L}$ \\
 \hline\hline
J2228.6-1636 &  AGN (unknown type) & 0.52 & $2.86\times10^{-12}$ & 12.31 \\
\hline
J1211.6+3901 &  BL Lac & 0.60 & $1.45\times10^{-12}$ & 10.79 \\
\hline
J2235.3-4836 &  FSRQ & 0.51& $1.05\times10^{-11}$ & 10.48 \\ 
\hline
J1435.9-8348 &  AGN (unknown type) & -- & $4.51\times10^{-12}$ & 9.29 \\ 
\hline
J0808.5+4950 &  FSRQ & 1.44 & $5.16\times10^{-12}$ & 8.84 \\ 
\hline
J1027.2+7427 &  AGN (unknown type) & 0.88 & $9.31\times10^{-12}$ & 8.33 \\
\hline
J1829.2-5813 &  FSRQ & 1.53 & $3.13\times10^{-11}$ & 8.04 \\ 
\hline
J0532.0-4827 &  BL Lac & -- & $4.49\times10^{-11}$ & 7.85 \\
\hline
J1401.2-0915 &  FSRQ & 0.67 & $3.13\times10^{-12}$ & 7.18 \\
\hline
J0928.4-0415 &  AGN (unknown type) & -- & $3.46\times10^{-12}$ & 7.11 \\
\hline
\end{tabular}
\caption{The 10 sources from the 4LAC catalog with the greatest evidence for neutrino emission, along with their redshift (when available) and gamma-ray flux as measured by Fermi~\cite{Fermi-LAT:2019pir}. None of these sources are statistically significant after accounting for an appropriate trials factor.}
\label{tbl:4lac_blazar_bright}
\end{table}

We begin by considering each of the 2860 sources in the 4LAC catalog independently (including both blazar and non-blazar AGN). In Table~\ref{tbl:4lac_blazar_bright}, we list the 10 of these sources that yielded the highest statistical significance in our analysis, along with their reported gamma-ray flux, redshift, and their source type as classified by the Fermi Collaboration~\cite{Fermi-LAT:2019pir}. While we cannot rule out the possibility that one or more of these sources is producing neutrinos, none is statistically significant after accounting for an appropriate trials factor. Furthermore, we have not identified anything about this collection of sources that sets them apart from other representative samples of sources within the 4LAC catalog; they do not appear to be systematically brighter (in gamma rays), nearby, or otherwise notable.

\begin{figure}[t]
\centering
\includegraphics[width=0.6\textwidth]{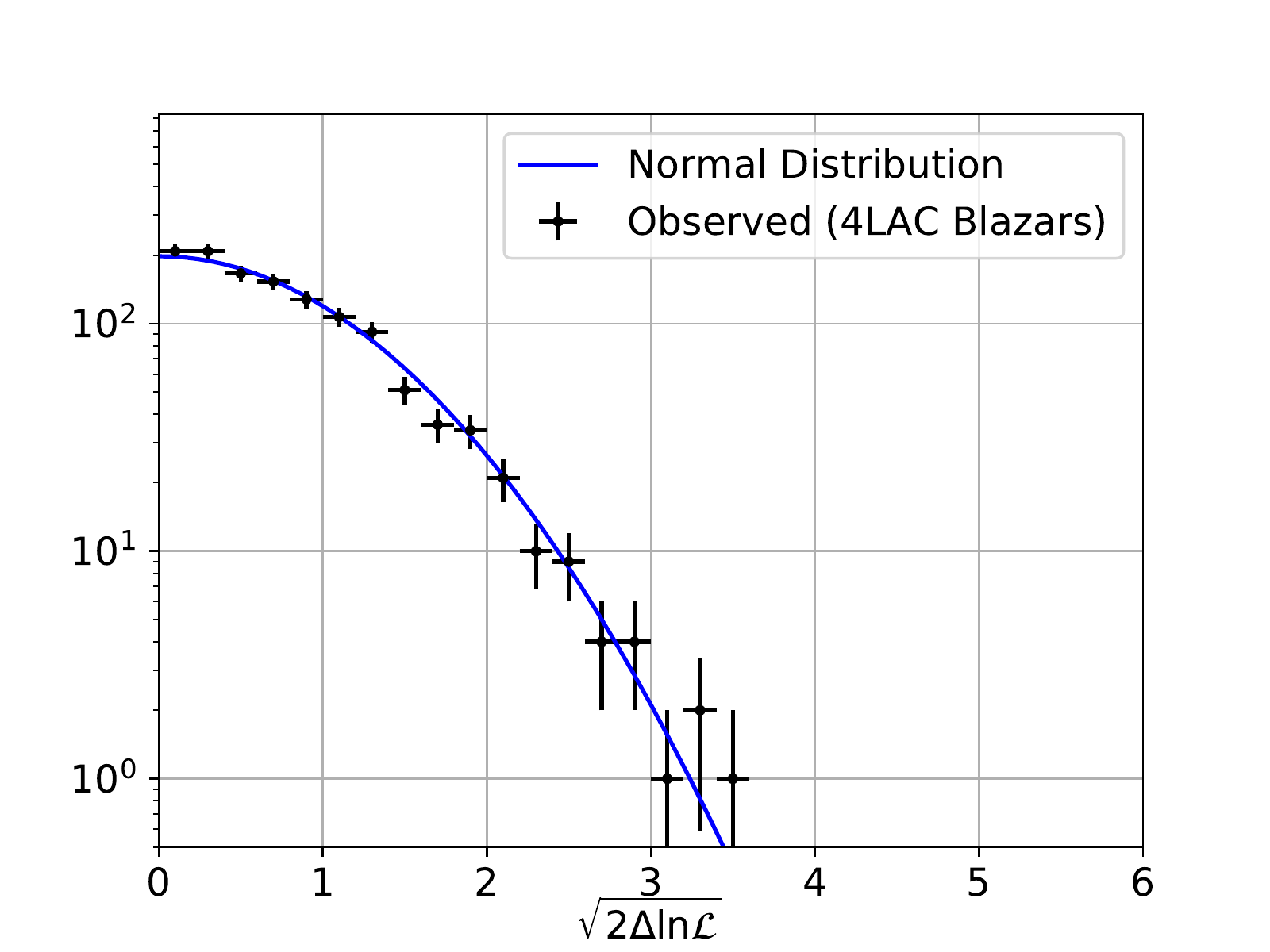}
\caption{The likelihood distribution in favor of a neutrino point source at the locations of 2796 blazars in the 4LAC catalog. The observed distribution is consistent with background and we identify no evidence of neutrino emission from this population of sources. Sky locations with $\Delta \ln \mathcal{L} < 0$, corresponding to a best fit with a negative point source flux, are not shown. The error bars represent the 68\% Poissonian confidence interval on each bin.}
\label{fig:blazar4lac}
\end{figure}

\begin{figure}[t]
\centering
\includegraphics[width=0.95\textwidth]{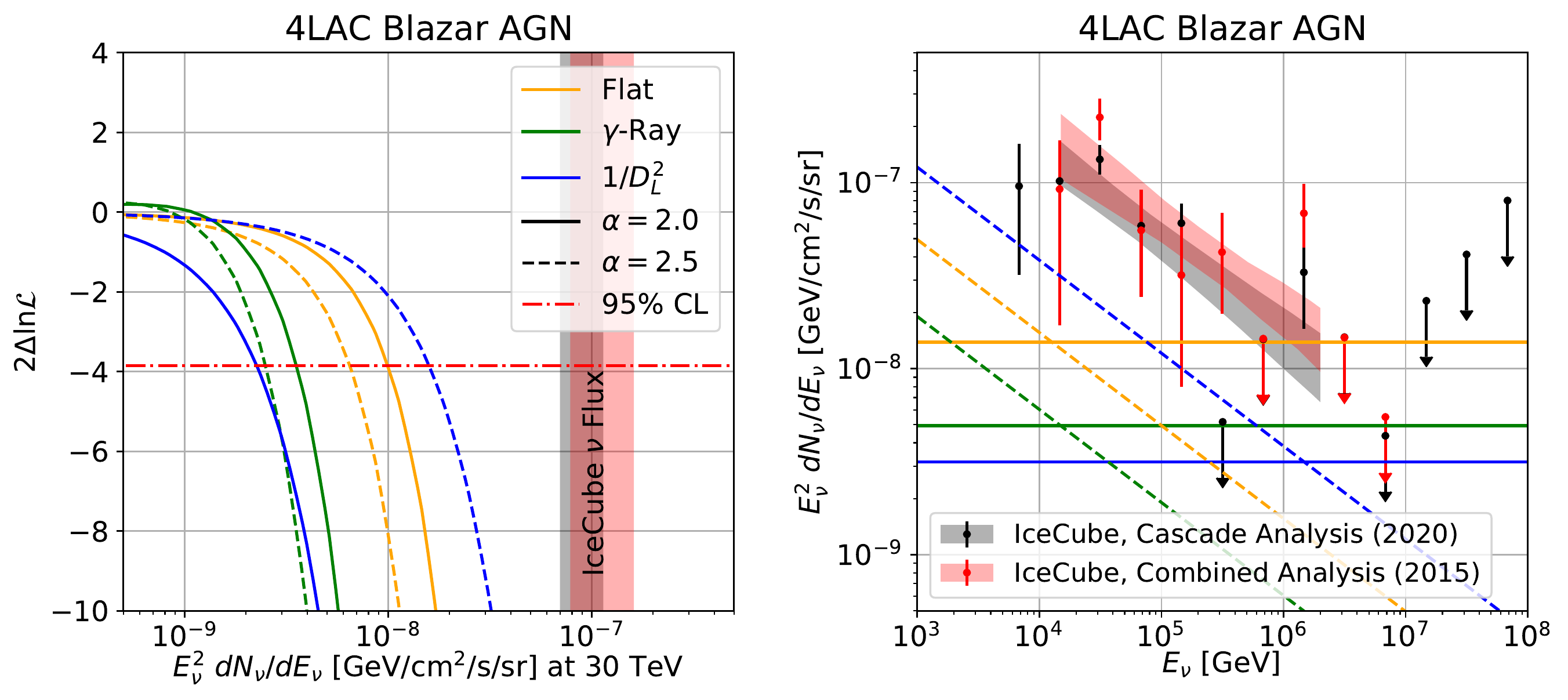} \\
\includegraphics[width=0.95\textwidth]{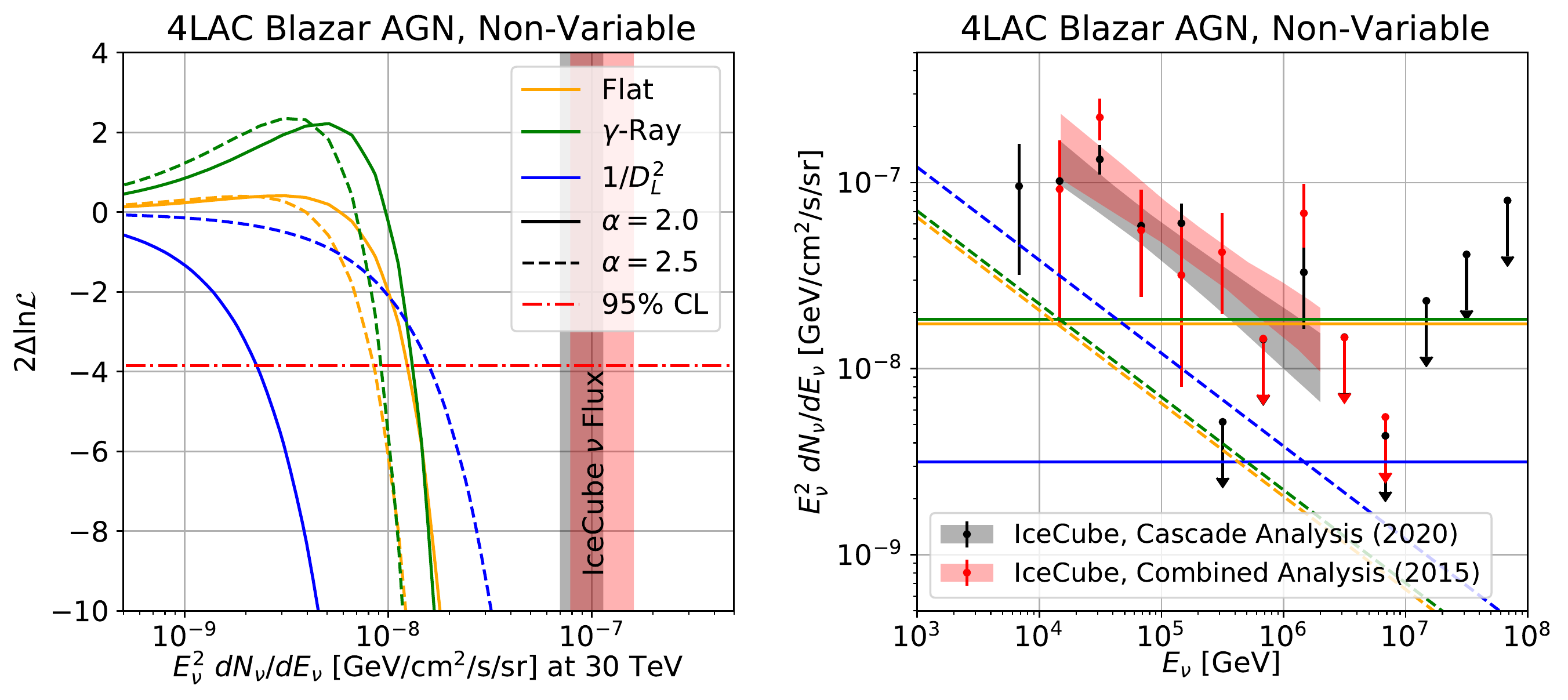}
\caption{In the left frames, we show the change to the log-likelihood as a function of the total, all-flavor neutrino emission from the blazars contained in the 4LAC catalog, for two choices of the neutrino spectral index, $\alpha$, and for the three flux weighting hypotheses described in Sec.~\ref{sect:methods}. We do not detect any statistically significant evidence for neutrino emission from this class of sources. In the upper frames, we show results based on all of the 2796 blazars contained in the 4LAC catalog, while in the lower frames we include only the 1674 of these blazars that do not exhibit statistically significant variability in their gamma-ray emission. In the right frames, we plot the 95\% confidence level upper limits on the total neutrino flux from these source populations, and compare these constraints to the diffuse neutrino flux as reported by the IceCube Collaboration~\cite{Aartsen:2015knd,Aartsen:2020aqd}. In the right frames, we have weakened the constraints by an appropriate completeness factor that accounts for the emission from blazars that are too distant or too gamma-ray faint to be included in the 4LAC catalog. From these results we conclude that no more than $\sim 15\%$ of IceCube's neutrino diffuse flux can originate from blazars.}
\label{fig:blazar4lac_limit}
\end{figure}

In Fig.~\ref{fig:blazar4lac}, we show the likelihood distribution of our analysis, for the 2796 sky locations associated with 4LAC blazars. As with the all-sky scan, we observe no significant evidence of any departure from Gaussian fluctuations. Next, we consider the joint probability of neutrino emission from this class of sources. In the left frames of Fig.~\ref{fig:blazar4lac_limit}, we show how the likelihood changes with the total neutrino flux from 4LAC blazars, for each of the three weighting hypotheses described in Sec.~\ref{sect:methods}, and for two choices of the spectral index (2.0 and 2.5). The value of $\alpha=2.5$ is motivated by the measured shape of IceCube's diffuse neutrino spectrum, whereas $\alpha=2.0$ is the value naively predicted from Fermi acceleration. For the results shown in the upper frames, we include all 2796 of the blazars in the 4LAC catalog. In the lower frames, we limit our analysis to the 1674 of these blazars (160 FSRQs, 700 BL Lacs, and 814 blazars of unknown type) that do not exhibit significant variability in their gamma-ray emission. More quantitatively, we include in the lower frames only those sources with a variability index below 18.48, as reported by the Fermi Collaboration~\cite{Fermi-LAT:2019yla}.\footnote{The variability index is defined as the difference in the log-likelihood between the flux fitted in each time interval and the average flux over the full catalog interval. Sources with a variability index greater than 18.48 are detected to be variable at the 99\% confidence level~\cite{Fermi-LAT:2019yla}.} In none of these cases do we obtain any statistically significant evidence for neutrino emission. We then proceed to place a 95\% confidence level upper limit (corresponding to $2\Delta \ln \mathcal{L} = -3.84$) on the total neutrino emission from this collection of sources. The limits we obtain are well below the diffuse flux reported by the IceCube Collaboration~\cite{Aartsen:2016xlq}, indicating that these blazars cannot produce a large fraction of the observed neutrinos.

In the right frames of Fig.~\ref{fig:blazar4lac_limit}, we show the upper limits that we have derived on the neutrino flux from blazars, and compare this to diffuse flux reported by the IceCube Collaboration. Here, we show the all-flavor diffuse flux as reported in IceCube's 2015 analysis of both cascades and muon tracks~\cite{Aartsen:2015knd}, and their more recent analysis of cascade events in six years of data~\cite{Aartsen:2020aqd}. In each case, we show the individual error bars associated with the measurement, as well as a shaded band, which represents the range of power-laws (between $\sim$\,20 TeV and $\sim$\,3 PeV) that is supported by the data.

In the left frame of this figure, these results specifically apply to the blazars contained in the 4LAC catalog. In order to translate this result to apply to the population of all blazars (including those that are too distant or gamma-ray faint to be included in the 4LAC), we weaken the limits in the right frames by a completeness correction factor of 1.4. Note that this quantity was originally calculated in Ref.~\cite{Aartsen:2016lir} in order to account for the fraction of the total gamma-ray emission from blazars that did not come from sources contained in the 2FGL catalog. Since the 4LAC catalog is significantly more complete than the 2FGL, a more accurate completeness factor would be less than 1.4 and closer to unity. In an effort to present a conservative constraint, however, we retain the value of 1.4 in our analysis. 

From the upper limits presented in this section, we conclude (at the 95\% confidence level) that blazars can produce no more than $\sim 15\%$ of the neutrino flux reported by IceCube. If we were to adopt the flat or gamma-ray scaling hypotheses, the constraints would be even more restrictive. These results are consistent with those presented by the IceCube Collaboration~\cite{Aartsen:2016lir} and in our previous analysis~\cite{Hooper:2018wyk}. We note that our analysis benefits from the use of the larger 4LAC catalog, which contains approximately $\sim60\%$ more sources than were utilized in our previous study. For completeness, we have also applied the above described procedure to those blazars categorized in the 4FGL as BL Lacs, or categorized as FSRQs. Again, we find no evidence in favor of neutrino emission from these source populations, allowing us to place limits on their contribution to IceCube's diffuse neutrino flux.

\section{Searching for Neutrinos from Non-Blazar AGN}

Although blazars cannot produce most of IceCube's diffuse neutrino flux, other types of AGN remain far more promising in this context. In particular, non-blazar AGN (those whose jets are not pointed in the direction of Earth) are less luminous and far more numerous than blazars, making it more difficult to constrain their contribution to the diffuse flux of high-energy neutrinos. Furthermore, it has been demonstrated that the isotropic gamma-ray background, as measured by the Fermi Gamma-Ray Space Telescope~\cite{Ackermann:2014usa}, is dominated by emission from unresolved non-blazar AGN~\cite{Hooper:2016gjy} (see also Refs.~\cite{Ajello:2015mfa,Cholis:2013ena,Fornasa:2015qua}). If the gamma-ray emission from these sources is generated through the interactions of cosmic-ray protons with gas, then they should also produce a spectrum of high-energy neutrinos that is similar to that measured by IceCube~\cite{Hooper:2016jls}. In contrast, if non-blazar AGN do {\it not} produce the bulk of IceCube's diffuse flux, then the sources of IceCube's neutrinos must produce their neutrino emission in environments that are optically thick to gamma rays, allowing the high-energy photons that are produced in conjunction with neutrinos through pion decay to be absorbed and thus not excessively contribute to the isotropic gamma-ray background.

In this section, we consider 65 non-blazar AGN, including the 63 contained in 4LAC catalog, as well as Centaurus B and 3C 411. Note that whereas the core and lobes of Centaurus A are listed as two sources in the 4LAC, we sum these gamma-ray fluxes and treat Centaurus A as a single source in our analysis. This collection of sources includes objects that Fermi classifies as compact steep-spectrum quasars, narrow line Seyfert 1 galaxies, radio galaxies, Seyfert galaxies, soft-spectrum radio quasars, and other non-blazar AGN of uncertain type. Of these 65 sources, 47 are non-variable, following the criteria described in Sec.~\ref{sect:4lac_blazar}. 

\begin{figure}[t]
\centering
\includegraphics[width=0.6\textwidth]{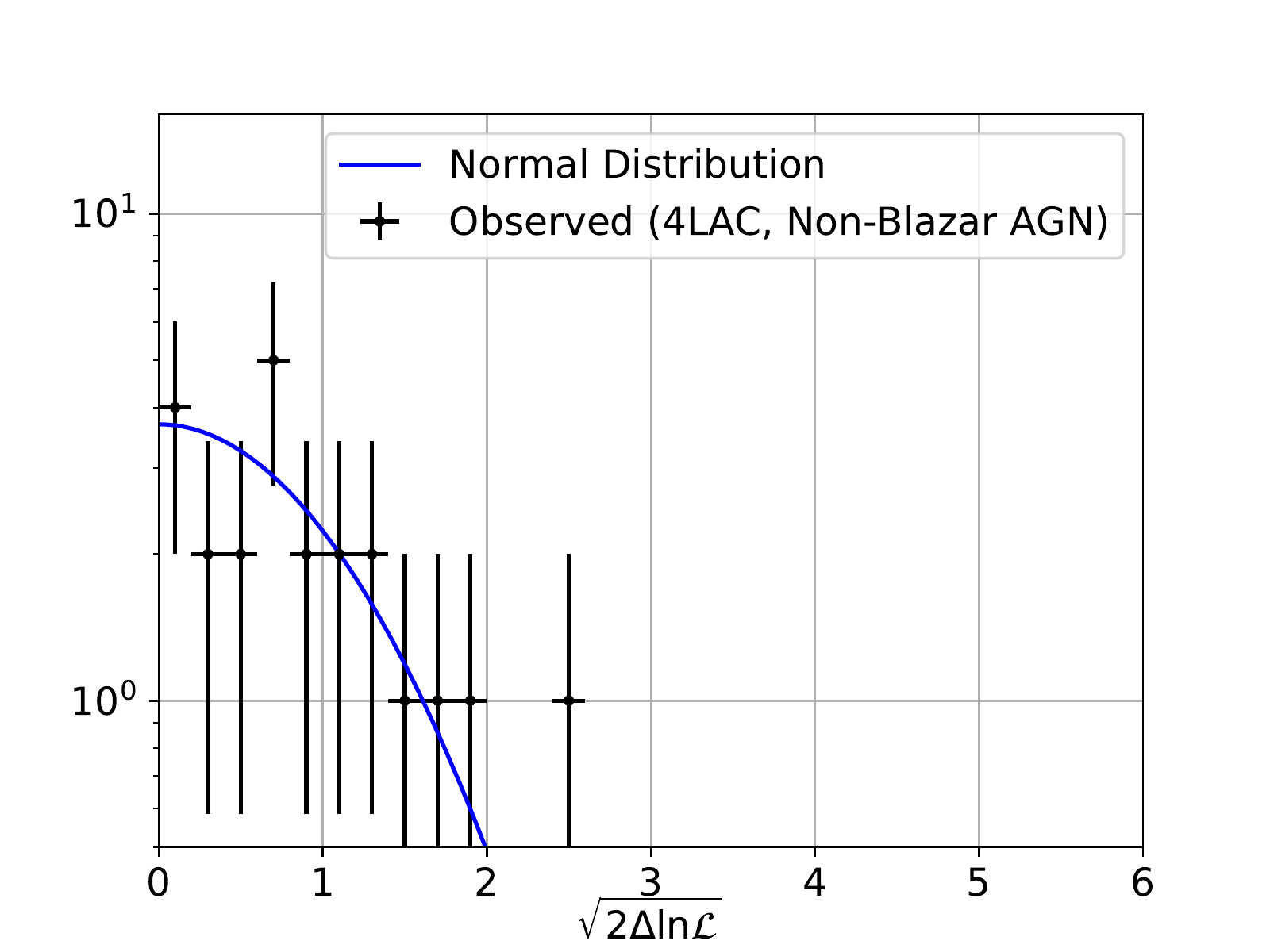}
\caption{The likelihood distribution in favor of a neutrino point source at the locations of 47 non-variable non-blazar AGN in the 4LAC catalog. The observed distribution is consistent with background and we identify no evidence of neutrino emission from this population of sources. Sky locations with $\Delta \ln \mathcal{L} < 0$, corresponding to a best fit with a negative point source flux, are not shown. The error bars represent the 68\% Poissonian confidence interval on each bin.} 
\label{fig:nonblazar4lac}
\end{figure}

\begin{figure}%[!h]
\centering
\includegraphics[width=0.95\textwidth]{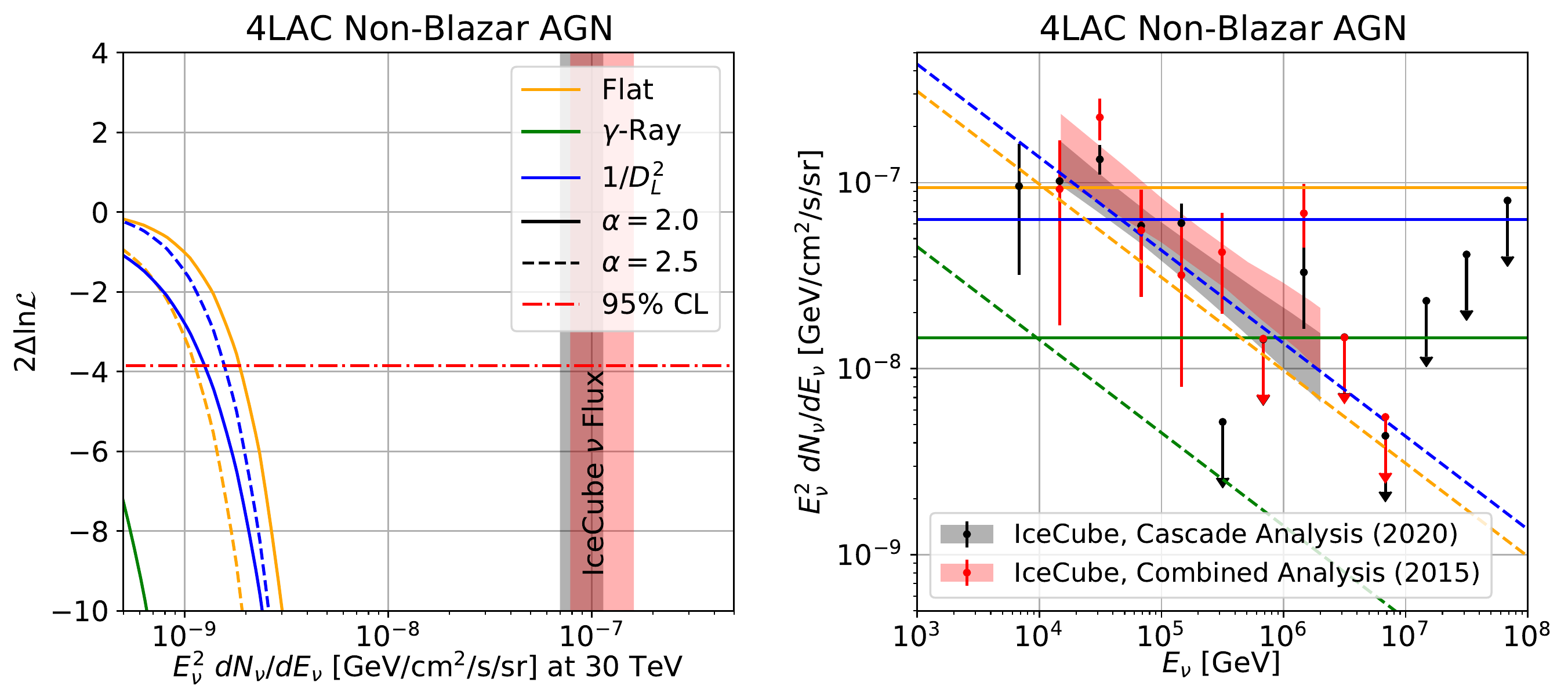}
\includegraphics[width=0.95\textwidth]{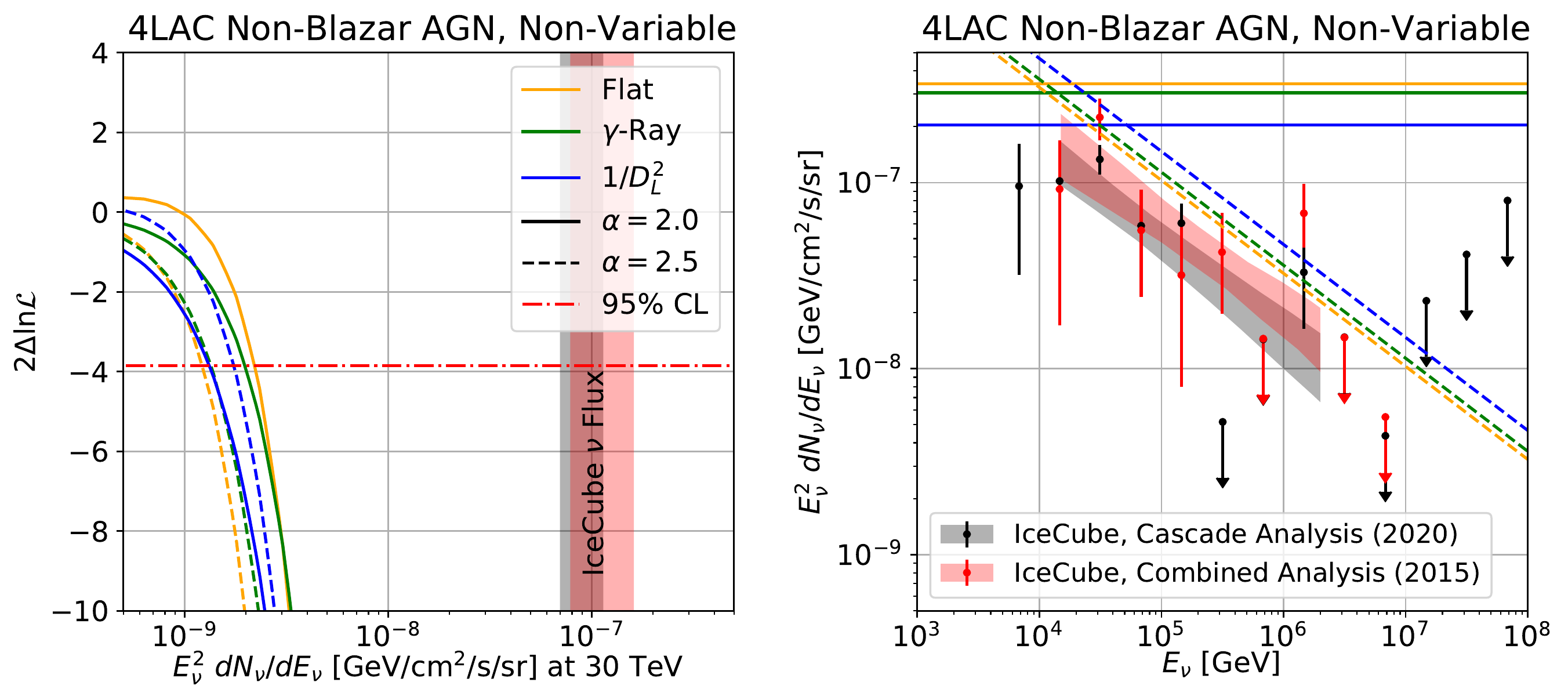}
\caption{In the left frames, we show the change to the log-likelihood as a function of the total, all-flavor neutrino emission from our sample of non-blazars AGN, for two choices of the neutrino spectral index, $\alpha$, and for the three flux weighting hypotheses described in Sec.~\ref{sect:methods}. We do not detect any statistically significant evidence for neutrino emission from this class of sources. In the upper frames, we show results based on all of the 65 non-blazar AGN contained in our sample, while in the lower frames we include only the 47 of these sources that do not exhibit statistically significant variability in their gamma-ray emission. In the right frames, we plot the 95\% confidence level upper limits on the total neutrino flux from these source populations, and compare these constraints to the diffuse neutrino flux as reported by the IceCube Collaboration~\cite{Aartsen:2015knd,Aartsen:2020aqd}. In the right frames, we have multiplied the constraints by an appropriate completeness factor that accounts for the emission from AGN that are too distant or too gamma-ray faint to be included in our sample. These results indicate that non-variable, non-blazar AGN could potentially generate the entirety of the diffuse neutrino flux reported by the IceCube Collaboration.}
\label{fig:nonblazar4lac_limit}
\end{figure}

In Fig.~\ref{fig:nonblazar4lac}, we plot the likelihood distribution for our sample of 47 non-variable, non-blazar AGN. Here, we have chosen to focus on the non-variable sources, as those AGN that exhibit a high degree of variability are generally throught to produce their gamma-ray emission primarily through leptonic processes. As with our results for the all-sky and blazar searches, this distribution is consistent with Gaussian fluctuations, with no statistically significant indication of neutrino emission.

The joint likelihood in favor of neutrino emission from non-blazar AGN is plotted in Fig.~\ref{fig:nonblazar4lac_limit}. In the upper frames, we include all 65 of the non-blazar AGN under consideration, while in the lower frames we limit our analysis to the 47 of these sources that do not exhibit significant variability in their gamma-ray emission. For no combination of spectral index and weighting hypotheses (see Sec.~\ref{sect:methods}) do we identify any evidence of neutrino emission. 

In the right frames of Fig.~\ref{fig:nonblazar4lac_limit}, we have again applied a completeness factor in order to account for those non-blazar AGN that are too distant or gamma-ray faint to be included in the 4LAC~\cite{Hooper:2018wyk}. Whereas the blazar completeness factor was not far above unity, a much larger fraction of the total gamma-ray emission from non-blazar AGN remains unresolved. Comparing the total gamma-ray flux from our sample of non-blazar AGN to the total contribution from all non-blazar AGN to the isotropic gamma-ray background~\cite{Hooper:2016gjy}, we find the appropriate completeness factor to be 50.6 (154.7) in the case of all (all non-variable) non-blazar AGN. 

The constraints shown in the upper frames of Fig.~\ref{fig:nonblazar4lac_limit} would appear to rule out the hypothesis that non-blazar AGN (including those that exhibit significant variability) produce the entirety of IceCube's reported signal. This conclusion is especially stark if we take the neutrino emission from a given source to be proportional to the observed gamma-ray flux. In this case, however, the limit is being driven in large part by a single source, NGC 1275. We excluded this source in our previous study~\cite{Hooper:2018wyk}, noting that the highly variable nature of NGC 1275's gamma-ray emission suggests that it is dominated by leptonic emission mechanisms. When this source is excluded from our analysis, the resulting constraint relaxes by a factor of approximately four, reducing the tension substantially.

After accounting for the lack of completeness in our catalog, we find that non-variable, non-blazar AGN could potentially generate the entirety of the diffuse neutrino flux reported by the IceCube Collaboration. It is encouraging to note, however, that these constraints are within a factor of $\sim$2-3 of the flux measured by IceCube. As IceCube accumulates and releases more data, and as gamma-ray catalogs of non-blazar AGN accumulate larger numbers of sources, we expect that it will become possible to definitively test the hypothesis that the majority of IceCube's signal originates from non-blazar AGN.

\section{Searching for Neutrinos from Starburst and Starforming Galaxies}

Starburst and other starforming galaxies are often discussed as a potential class of sources for IceCube's diffuse neutrino flux. Although these sources cannot produce the entirety of IceCube's signal without exceeding the measured intensity of the isotropic gamma-ray background~\cite{Bechtol:2015uqb}, it remains entirely plausible that they could generate a non-negligible fraction (up to $\sim$10\%) of the neutrinos observed by IceCube. As they are even more numerous and less luminous than non-blazar AGN, it is expected to be very difficult to detect neutrinos from individual starforming galaxies.

\begin{figure}%[!h]
\centering
\includegraphics[width=0.95\textwidth]{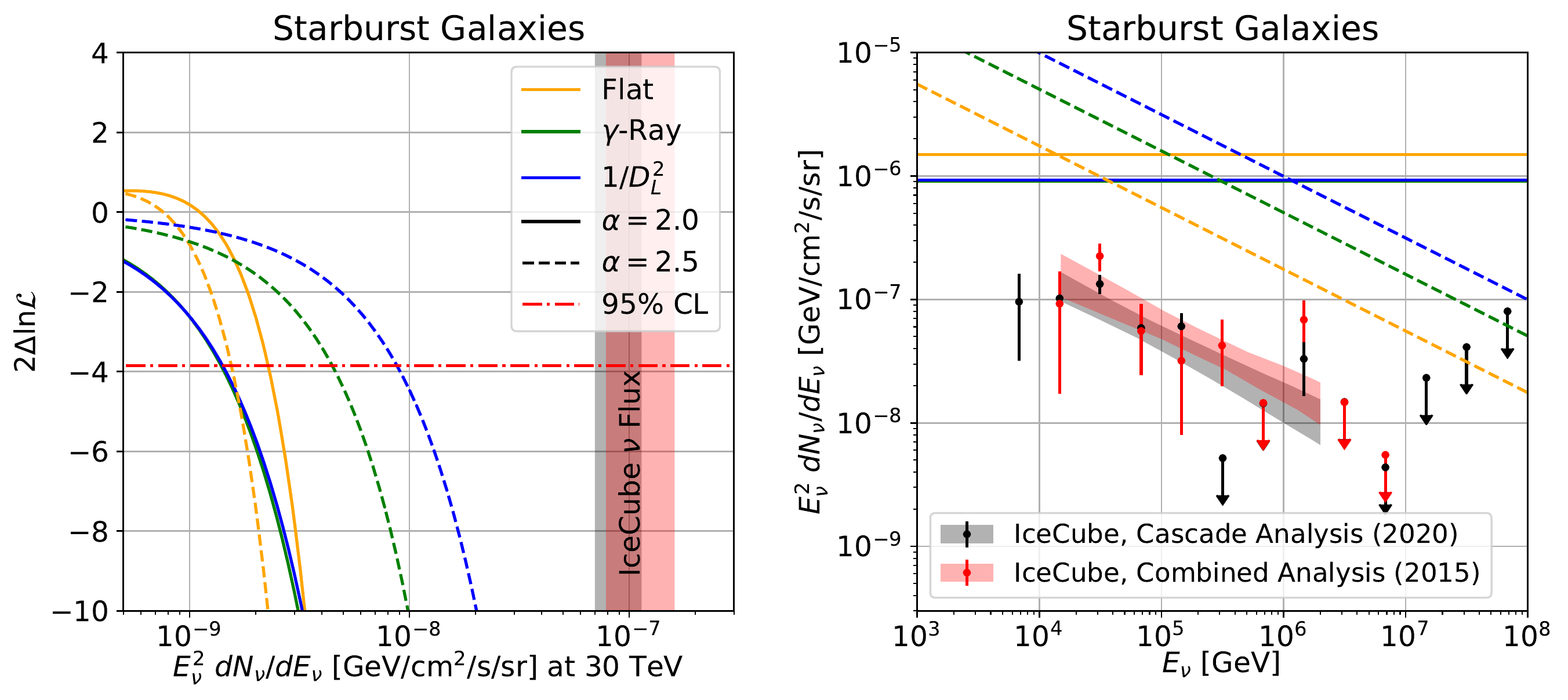}
\caption{In the left frame, we show the change to the log-likelihood as a function of the total, all-flavor neutrino emission from our sample of 45 nearby radio- and infrared-bright starburst galaxies~\cite{Lunardini:2019zcf}, for two choices of the neutrino spectral index, $\alpha$, and for the three flux weighting hypotheses described in Sec.~\ref{sect:methods}. We do not detect any statistically significant evidence for neutrino emission from this class of sources. In the right frame, we plot the 95\% confidence level upper limits on the total neutrino flux from this source population, and compare these constraints to the diffuse neutrino flux as reported by the IceCube Collaboration~\cite{Aartsen:2015knd,Aartsen:2020aqd}. In the right frame, we have multiplied the constraints by an appropriate completeness factor that accounts for the emission from starburst and starforming galaxies that are too distant or too gamma-ray faint to be included in our sample.}
\label{fig:sbg_limit}
\end{figure}

For the analysis performed in this section, we make use of a catalog of 45 nearby radio- and infrared-bright starburst galaxies, as described in Ref.~\cite{Lunardini:2019zcf}. Once again, we find no evidence of neutrino emission from this class of sources. In the right frame of Fig.~\ref{fig:sbg_limit}, we have applied a completeness factor of 650, which is significantly larger than in the case of blazar or non-blazar AGN due to the significantly lower luminosities of these sources. Once again, we estimate this factor by comparing the intensity of the isotropic gamma-ray background to the combined gamma-ray emission from our sample of 45 sources. We determine this later quantity using the reported measurements of the far-infrared emission from these sources, which we relate to their gamma-ray emission by applying the empirical correlation described in Refs.~\cite{Linden:2016fdd,Ackermann_2012,Rojas-Bravo:2016val}. We note that there are significant uncertainties associated with the determination of this correction factor for starburst and starforming galaxies, and the constraints presented in the right frame of Fig.~\ref{fig:sbg_limit} could plausibly be inaccurate at the level of a factor of $\sim$\,2-3.

\section{Summary and Conclusions}

The origin of IceCube's diffuse flux of high-energy neutrinos remains one of the most interesting and important open questions in the field of high-energy astrophysics. In an effort to shed light on this mystery, we have used 3 years of publicly available IceCube data to measure or constrain the fraction of IceCube's flux that originates from blazars, non-blazar AGN, and starforming galaxies. 

Our analysis did not identify any statistically significant neutrino emission from any of the source classes under consideration. Instead, our results force us to conclude that no more than 15\% of IceCube's diffuse high-energy neutrino flux can originate from blazars. In contrast, it remains possible that non-blazar AGN could produce the entirety of the neutrino flux observed by IceCube. We expect such a scenario to be testable in the relatively near future, as a result of additional IceCube data and increasingly complete catalogs of gamma-ray AGN. Our constraints on starburst and other starforming galaxies remain quite weak, and we cannot significantly test the hypothesis that such sources contribute significantly to IceCube's signal.

\acknowledgments

DS and AGV are supported by a Cottrell Scholar Award, a program of Research Corporation for Science Advancement, and a NASA Roman Technology Fellowship, award number 80NSSC19K0298. DH is supported by the Fermi Research Alliance under Contract No.~DE-AC02-07CH11359 with the U.S. Department of Energy, Office of High Energy Physics.

\bibliography{references}
\bibliographystyle{JHEP}

\end{document}